# Absence of magnetic order in epitaxial RuO$_2$ revealed by X-ray linear dichroism

Siyu Wang[1#], Chao Wang[1#], Yanan Yuan[1], Jiangxiao Li[1], Fangfang Pei[1], Daxiang Liu[1], Chunyu Qin[1], Jiefeng Cao[2], Yamei Wang[2], Tianye Wang[3], Jiayu Liu[1], Jieun Lee[3], Guanhua Zhang[4], Christoph Klewe[5], Chenchao Yu[6], Fan Zhang[7], Dongsheng Song[3], Kai Chen[1], Weisheng Zhao[5], Dawei Shen[1], Ziqiang Qiu[3], Mengmeng Yang[6*], Bin Hong[7*], and Qian Li[1*]

[1]*School of Nuclear Science and Technology and National Synchrotron Radiation Laboratory, University of Science and Technology of China, Hefei, Anhui 230029, China*
[2]*Shanghai Synchrotron Radiation Facility, Shanghai Advanced Research Institute, Chinese Academy of Sciences, Shanghai 201204, China*
[3]*Physics department, University of California at Berkeley, Berkeley, California 94720, USA*
[4]*State Key Laboratory of Molecular Reaction Dynamics, Dalian Institute of Chemical Physics, Chinese Academy of Sciences, Dalian, China*
[5]*Advanced Light Source, Lawrence Berkeley National Laboratory, Berkeley, California 94720, USA*
[6]*Institute of Physical Science and Information Technology, Anhui University, Hefei, Anhui 230601, China*
[7]*Nanoelectronics Science and Technology Center, Hefei Innovation Research Institute, Beihang University, Hefei 230013, China*

Recently, the topic of altermagnetism has attracted tremendous attention and RuO$_2$ have been demonstrated to be one of the most promising altermagnetic candidates. However, disputes still remain on the existence of magnetic order in RuO$_2$. Here in this work, we employ X-ray linear dichroism (XLD), a widely utilized technique for characterizing antiferromagnets, in conjunction with photoemission electron microscopy and multiple scattering calculation to provide clear evidence of the absence of magnetic order in epitaxial RuO$_2$ films. The observed XLD signal is nearly invariant with temperature and independent on cooling field direction, in stark contrast to the substantial magnetic order-related XLD signal predicted by multiple scattering calculation. This finding strongly suggests a nonmagnetic origin for RuO$_2$. Furthermore, we observed significantly distinct XLD signals at the Ru M$_3$ and O K edges in RuO$_2$ films grown on TiO$_2$ substrate with different surface orientations, which can be attributed to the low-symmetry crystal field. These results unequivocally demonstrate the absence of magnetic order in RuO$_2$ and establishes XLD measurement as a robust technique for probing the low-symmetry magnetic materials.

Altermagnets, recently identified as a distinct class of magnetic materials characterized by zero net magnetization, have garnered significant attention due to their intriguing properties, which differ markedly from those of conventional antiferromagnets (AFMs)[1-5]. Unlike conventional AFMs, which exhibit spin-degenerate bands governed by intersublattice symmetries such as inversion or translation, altermagnets feature spin-split energy bands with opposite-spin sublattices connected by real-space rotational symmetry. This alternating spin polarized electronic band structure was first predicated theoretically[2-4] and has been recently corroborated by experimental spectroscopic studies[6-8]. RuO$_2$, a metallic compensated altermagnet candidate with rutile structure, has exhibited several unconventional phenomena, including an anomalous Hall effect[4,9,10], as well as novel crystalline surface- and Néel vector-dependent spin-charge[11-13] and charge-spin[14-16] conversion effects arising from its spin-splitting properties. However, the interpretation of this altermagnetic spin-splitting effect is complicated by its potential overlap with the anisotropic spin Hall effect [17,18], and the presence of significant electrical anisotropy [19], raising questions about the existence of a distinct altermagnetic spin-splitting effect. The anomalous Hall effect was recently found hard to be distinguished from the ordinary Hall contributions[20]. Furthermore, consensus on the presence and nature of magnetic order in RuO$_2$ remains elusive.

Previous resonant x-ray scattering[21] and neutron diffraction[22] experiments reported AFM order in RuO$_2$ persisting above 300 K, with a relatively small estimated magnetic moment value (~0.05 μ$_B$/Ru). However, the proposed magnetic origin has been challenged by subsequent studies due to existence of phase shift errors[23] and multiple scattering artifacts [24,25] in previous work, casting doubt on its validity. Theoretical investigations suggest that itinerant antiferromagnetism in RuO$_2$ could arise from the instability due to Fermi surface nesting, potentially leading to a spin density wave (SDW) state[22,26]. Nevertheless, there is ongoing debate regarding whether this SDW is commensurate[21] or incommensurate[27]. Notably, no evidence of phase transition to AFM phase was observed in the heat capacity[28,29] or the magnetic susceptibility [30,31] measurements of RuO$_2$. Previous nuclear magnetic resonance revealed no detection of magnetic hyperfine field on Ru[32] and the recent broadband infrared spectroscopy on RuO$_2$ find the result well described by a paramagnetic metal[33]. Recent muon spin rotation/relaxation[24,34] experiments detected an exceedingly small magnetic moment (~$10^{-4}$ $\mu_B$/Ru), further casting doubt on the presence of AFM order in RuO$_2$. Additionally, conflicting results have emerged from angle-resolved photoemission spectroscopy (ARPES) studies, with one report indicating an absence of band-splitting[35], and another claiming evidence for altermagnetism in RuO$_2$ [36]. A recent theory work proposed that RuO$_2$ is intrinsically nonmagnetic, attributing the reported AFM order to hole doping induced by Ru vacancies[37]. Furthermore, epitaxial strain of RuO$_2$ film, influenced by substrate lattice mismatch, has been shown to significantly affect the crystal symmetry[38] and modify the orbital/electronic structure near the Fermi level[39].

These conflicting observations highlight the urgent need to resolve the nature of AFM order in RuO$_2$ films through direct experimental evidence. X-ray linear dichroism (XLD) is a powerful technique widely used for detecting compensated AFM order in thin films. As one of the few methods exhibiting element-specific and structural sensitivity, capable of probing both the crystal field and magnetic order in AFM films, XLD has successfully elucidate the AFM spin structures of materials such as LaFeO$_3$[40], NiO[41-43], CoO[44-48], BiFeO$_3$[49], van der Waal antiferromagnets such as CrSBr[50], and etc. In this work, we employed XLD to investigate crystal field- and magnetic order-related XLD signal in epitaxial RuO$_2$ films on TiO$_2$ substrate with varying crystal orientations. The observed near-temperature-invariant and cooling field direction-independent XLD effects, combined with spatially resolved photoemission electron microscopy (PEEM) results and multiple scattering calculations, strongly indicate the absence of magnetic order in RuO$_2$.

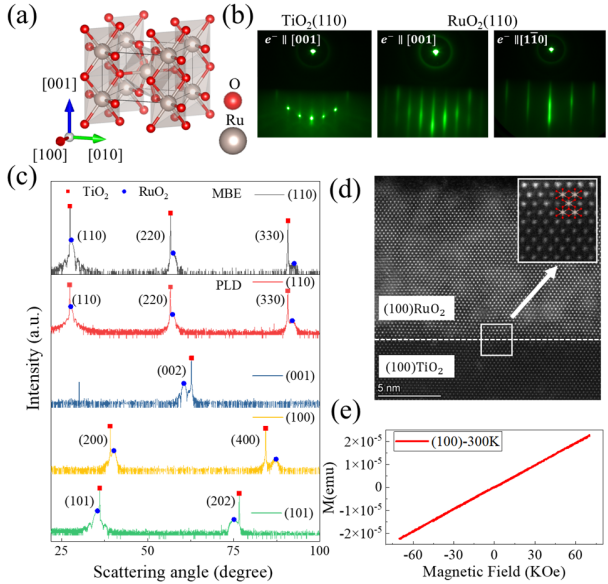

FIG. 1: (a) Schematic drawing of crystal structure in RuO$_2$. (b) RHEED patterns of MBE-grown TiO$_2$(110) substrate and RuO$_2$(110) film. (c) XRD characterization of RuO$_2$ films on TiO$_2$ substrates grown by both MBE and PLD methods with different crystal surfaces of (110), (001), (100), and (101). Red and blue dots represent the diffraction peaks corresponding to TiO$_2$ substrate and RuO$_2$ film, respectively. (d) High-resolution transmission electron microscopy (HRTEM) image of epitaxial RuO$_2$ film on TiO$_2$(100), showing excellent film quality. (e) VSM measurements of the RuO$_2$(100)/TiO$_2$ samples at room temperature.

RuO$_2$ is a rutile oxide with a crystallographic structure described by the space group of $P4_2/mnm$[36], where Ru atoms occupy the center of stretched oxygen octahedrons, as depicted in Fig. 1(a). Since the magnetic properties of RuO$_2$ are known to be highly sensitive on the growth conditions, we prepared epitaxial RuO$_2$ thin films on TiO$_2$ substrates using both pulsed laser deposition (PLD) and molecular beam epitaxy (MBE), enabling comparative XLD measurements. Detailed growth parameters are provided in the Supplemental Material (Supplementary Part 1[51]). The high crystalline quality of our films is evidenced by the sharp reflection high-energy electron diffraction (RHEED) patterns [Fig. 1(b)] and well-defined X-ray diffraction (XRD) peaks for films grown on (110), (001), (100), and (101) TiO$_2$ surfaces [Fig. 1(c)]. The cross-sectional high-angle annular dark-field scanning transmission electron microscopy (HAADF-STEM) images reveal an atomically sharp film-substrate interface and preserved crystallographic symmetry [Fig. 1(d), shown for the (100) plane along the [001] direction]. Complementary x-ray reflectivity, atomic force microscopy measurements, and reciprocal space maps, combined with lattice constant analysis (Supplementary Fig. S1, S2 and Table S1[51]), confirm low surface roughness and minimal lattice mismatch. Vibration sample magnetometry (VSM) measurement reveal a strictly linear magnetization-field dependence with negligible magnetic moment, consistent with previous reports[21,22,34]. These results collectively provide compelling evidence for the absence of ferromagnetic order in RuO$_2$ film.

Given that the reorientation of the Néel vector at room temperature requires an exceptionally strong magnetic field exceeding 50 T[9,10], in-situ X-ray magnetic linear dichroism (XMLD) measurements are challenging to perform under such conditions. Instead, the XLD measurements can be conducted by altering the X-ray linear polarization with X-ray at normal incidence, as illustrated in Fig. 2(a). It is important to note that the XLD effect may originate from both crystal field effects and the magnetic order contributions [41-48]. Therefore, careful differentiation between these two effects is essential before attributing the XLD signal soley to magnetic order. This distinction is particularly critical in the case of RuO$_2$, which exhibits a low-symmetry rutile crystal structure that can significantly influence the crystal field contributions to the XLD effect.

Fig. 2(b) illustrates schematic representations of various crystal surfaces. The (110), (100) and (101) surfaces exhibit significantly different lattice constants and atomical environment along the horizontal and vertical axes, whereas the (001) surface displays identical lattice constants and atomic environments in both directions. A summary of crystal orientations corresponding to the horizontal ($\varphi = 0°$) and vertical ($\varphi = 90°$) axes for different surfaces is provided in the table shown in Fig 2(c). The XLD measurements were conducted at BL07U of Shanghai synchrotron radiation facility (SSRF)[59,60] and BL4.0.2 of Advanced light source (ALS). XLD effect were investigated at both the Ru $M_3$ edge (3p→4d, Fig. 2(d)) and the O K edge (1s→2p, Fig. 2(e)). At the Ru $M_3$ edge, the X-ray absorption spectra (XAS) revealed multiple peaks at approximately 461.8 eV and 463.8 eV, characteristic of the multiplet splitting of Ru 4d energy levels [61,62]. The XLD signal, defined as the difference between XAS results with X-ray polarizations aligned along $\varphi = 0°$ and $\varphi = 90°$, is presented for the $RuO_2$(110) surface in the second panel of Fig. 2(d). The XLD signal for the (110) surface exhibited a prominent negative peak at ~461.8 eV with an intensity of ~−11.5% and a positive peak at ~465.3 eV with an intensity of ~4.7%. A similar spectra shape, but with a slightly differing magnitudes, was observed for the (110) and (101) surfaces. In contrast, the (001) surface displayed negligible XLD signals, suggesting minimal anisotropy in their electronic structures. The $RuO_2$ films used in this study are sufficiently thick to suppress the Ti L-edge signal from the substrate (Supplementary Fig. S3[51]).

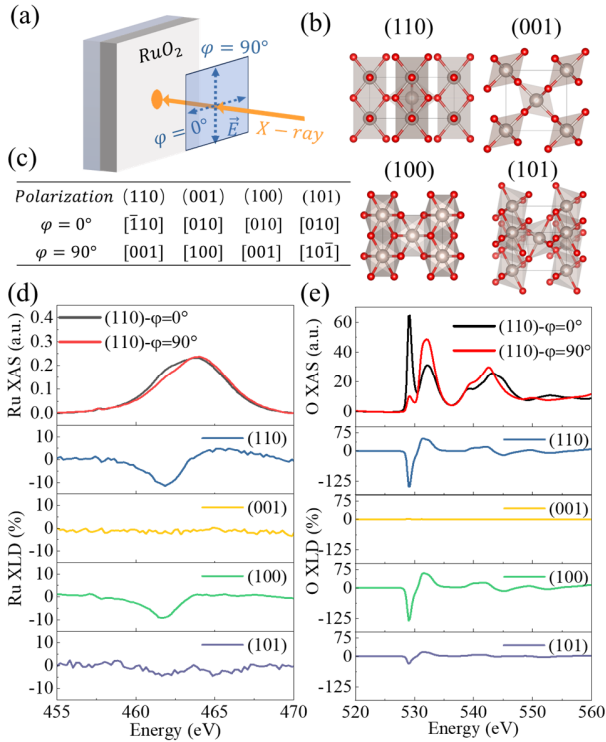

FIG. 2: (a) Schematic drawing of the x-ray linear dichroism measurement with switched x-ray polarizations, where the angle $\varphi$ represent the angle between the horizontal crystal axis and the linear polarization direction. (b) Schematic drawing of crystal lattices for $RuO_2$ (110), (001), (100), and (101) surfaces with grey atoms representing Ru lattice sites and red atoms representing O lattices sites. (c) a table listing the orientations of crystal axes parallel to the linear polarization direction of x-ray for the situations of $\varphi = 0°$ and $\varphi = 90°$. Typical X-ray absorption spectra at (d) Ru $M_3$ edge and (e) O K edge from $RuO_2$(110) with $\varphi = 0°$ and $\varphi = 90°$, respectively, and XLD signal measured at (d) Ru $M_3$ edge and (e) O K edge from $RuO_2$(110), (001), (100) and (101) surfaces deposited by PLD. All measurements were performed at room temperature.

Despite the delocalized nature of Ru 4d states, their hybridization with O 2p states offers an alternative approach for characterizing Ru 4d states through the O K-edge spectrum[42,62,63]. In the lower-symmetry octahedral ($O_h$) crystal field, $Ru^{4+}$ ($4d^4$) states are split into $e_g$ set (comprising xy and $3z^2-r^2$ orbitals) and $t_{2g}$ set (comprising $x^2-y^2$, xz, and yz orbitals)[62]. As shown in Fig. 2(e), the O K-edge spectrum prominently displays two peaks at ~529.0 eV and ~532.0 eV, which can be attributed to the coupling of O 2p states with Ru 4d $t_{2g}$ and $e_g$ state, respectively. The peaks in the energy range of 535 eV to 550 eV are associated to hybridized 2p states with Ru 5sp bands[63,64]. The energy separation between the $t_{2g}$ and $e_g$ state is approximately 3.0 eV, providing an estimate of the crystal field splitting that consistent with previous reports[62]. X-ray polarization direction significantly influenced the XAS intensities to the $t_{2g}$ state and $e_g$ states, yielding an XLD signal of ~−145.2% at ~529.0 eV and ~51.5% at ~531.4.0 eV for $RuO_2$(110) surface. The $RuO_2$(100) and $RuO_2$(101) surfaces exhibited similar spectra shapes but with reduced XLD signal magnitudes, while the $RuO_2$(001) surfaces showed negligible XLD signals. These observations align with the XLD results obtained at the Ru $M_3$ edge. Furthermore, the crystal field splitting remains nearly unchanged across different $RuO_2$ crystalline orientations (Supplementary Fig. S4[51]). Notably, the magnitude of the XLD signal at the O K edge is an order of magnitude larger than that at Ru $M_3$ edge, highlighting the O K edge as a highly sensitive probe for detecting anisotropic crystal fields and magnetic order within the $RuO_2$ system.

Previous studies investigating the potential existence of magnetic order in $RuO_2$[4,9-19,21,22] suggest that the spin direction aligns along [001] axis, which coincides with the central symmetry axis of the $RuO_2$ crystal structure. Consequently, the magnetic order-related XLD signal is inherently coupled with and indistinguishable from the crystal field-related XLD signal across different crystal surfaces, making it challenging to discern the origin of XLD effect in Fig. 2. To isolate the magnetic order-related XLD signal in $RuO_2$, a widely adopted approach involves measuring the temperature-dependence of the XLD signal across the AFM to paramagnet (PM) phase transition [41,43,46,48,50]. Above the Néel temperature of an AFM, the contribution from magnetic order diminishes, leaving only the crystal field contribution to the XLD signal.

Following this approach, we measured the XLD signal of RuO$_2$(110) at both the Ru M$_3$-edge [Fig. 3(a)] and O K-edge [Fig. 3(b)] over a temperature range of 310 K to 450 K, covering the critical temperature range of AFM to PM phase transition[21,34]. The XLD signal at both Ru M$_3$ edge (461.8 eV) and O K edge (529.0 eV) exhibits a weak temperature dependence in the range of 310 K to 450 K, with a standard deviation of XLD signal ~0.25% and ~1.9% for Ru and O, respectively. Similar temperature-insensitive XLD behavior was observed for RuO$_2$ films deposited by PLD including different surfaces of RuO$_2$(110), RuO$_2$(100), and RuO$_2$(001), and different measurement geometries of comparing normal and grazing incidence geometries with fixed X-ray polarization (Supplementary Fig. S5[51]). These findings suggest that the XLD signal in RuO$_2$ is dominated by crystal field effects rather than magnetic order.

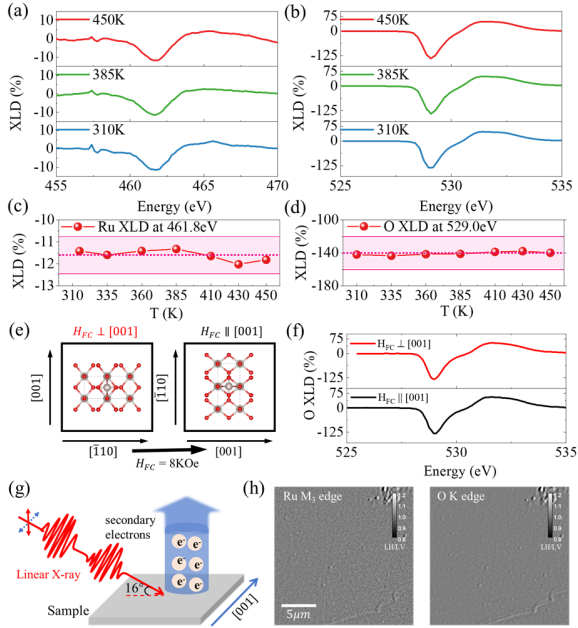

FIG. 3: Temperature-dependent XLD results from RuO$_2$(110) at (a) Ru M$_3$ edge and (b) O K edge covering temperature range of 310 K to 450 K. Summary of the temperature-dependent XLD signal at (c) 461.8 eV and (d) 529.0 eV, where the red dashed line denotes the average XLD value, and the red region represents the range of calculated XLD variation between magnetic and nonmagnetic configurations from Fig 4. (e) O XLD signal measured at room temperature from two samples of RuO$_2$(110) after field cooling from 200°C to RT with H$_{FC}$=8 KOe, H$_{FC}$//[001] and H$_{FC}$//[$\bar{1}$10] (i.e. denoted as H$_{FC}$⊥ [001]), respectively.(g) Schematic drawing of PEEM imaging geometry with two perpendicular linear polarizations. (h) XMLD domain images taken at the Ru M$_3$ edge and O K edge at room temperature. The field view is 20 μm.

An additional effective method to differentiate the influence of magnetic order from crystal field effects on the XLD signal involves employing a field-cooling procedure to reorient the direction of Néel vector in RuO$_2$ film. This technique has been successfully applied in systems such as Py/RuO$_2$ system by utilizing exchange coupling at the interface [13,16], as well as in single AFM system (e.g. CoO, NiO and MnTe) through spin-flop transitions during the field cooling process[43,47,65]. Notably, the magnetic field required to reorient AFM spins is significantly reduced at high temperatures. In this study, two RuO$_2$(110) samples, cut from the same parent sample, were subjected to field-cooling procedures with H$_{FC}$=8 KOe, cooled from 200°C to room temperature along RuO$_2$[001] and [$\bar{1}$10] directions, respectively [Fig. 3(e)]. Unlike previous observations in rotatable CoO[44-46,48] and NiO films [42], where reorientation of AFM spins resulted in reversed XLD signals, the two RuO$_2$ samples exhibited nearly identical XLD signals after field cooling along orthogonal directions, as shown in Fig. 3(f). The XLD signal at ~529.0 eV differed by only ~0.82%, indicating the absence of a XLD signal corresponding to reorientation of RuO$_2$ Néel vector or the repopulation of multidomain structures during field-cooling process. These results strongly suggest that field cooling does not induce Néel vector reorientation in RuO$_2$, further supporting the absence of magnetic order-related contributions to the XLD signal in this system.

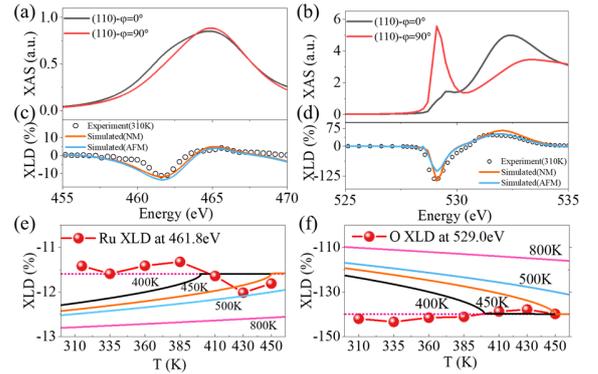

FIG. 4: Multiple scattering calculation result of RuO$_2$ XAS at (a) Ru M$_3$ edge and (b) O K edge for the situations of $\varphi = 0°$ (i.e., E//RuO$_2$[$\bar{1}$10]) and $\varphi = 90°$ (i.e., E//RuO$_2$[001]) from RuO$_2$(110) surface. Comparisons of calculated XLD signal at (c) Ru M$_3$ edge and (d) O K edge supposing non-magnetic (light blue lines) and magnetic signature of RuO$_2$ (dark blue line) with experimental results (black circles). Comparison of experimental temperature-dependent XLD signal at (e) 461.8 eV and (f) 529.0 eV with calculated XLD(T) curves supposing magnetic RuO$_2$ with different T$_N$ values following the scaling law with $\beta = 0.3$.

To elucidate the specific contribution of magnetic order to the XLD signal, we conducted theoretical calculations based on multiple scattering. The Ru M$_3$-edge and O K-edge XAS and XLD spectra were simulated using the multiple scattering method in Green's formalism, as implemented in the FDMNES software[66]. Multiple scattering paths were summed within a spherical cluster of radius 9 Å as determined by convergence tests. Quadrupolar transitions and spin-orbit coupling were included in all calculations to ensure accuracy. Energy-dependent linewidth broadening were considered for the convolution of simulated XAS spectra (Supplementary Part 2[51]) and core level width was set according to previous report[67]. For the case of magnetic RuO$_2$, a self-consistent magnetic moment of 0.9 μ$_B$/Ru was applied, consistent with interpretations of earlier experiments[9,10,21,22]. This approach enabled the theoretical separation of the contributions from magnetic order and crystal field effects to the observed XLD signal.

Our calculation reveals that the simulated XAS at Ru M$_3$-edge (Fig. 4(a)) and O-K edge (Fig. 4(b)) for nonmagnetic RuO$_2$ agree closely with the experimental results, whereas the magnetic configuration show significant deviations (Supplementary Fig. S6[51]). Figure 4(c) and (d) compare the experimental XLD signals with calculations for two scenarios: nonmagnetic RuO$_2$ and magnetic RuO$_2$ with the Néel vector aligned along [001] at Ru M$_3$ edge and O K-edge, respectively. At the Ru M$_3$ edge (461.8 eV), the XLD intensity for nonmagnetic RuO$_2$(110) is $\sim-11.8\%$, while the magnetic configuration yields a stronger XLD signal of $\sim-13.5\%$. At the O K edge (529.0 eV), the nonmagnetic case exhibits a much stronger XLD signal ($\sim-145.8\%$) compared to the magnetic case ($\sim-105.4\%$) between x-ray polarization parallel to RuO$_2$[$\bar{1}$10] ($\varphi = 0°$) and RuO$_2$ [001] ($\varphi = 90°$). The results align well with the experimental XLD signal of $\sim-11.6\%$ at 461.8 eV and $\sim-140\%$ at 529.0 eV, further supporting the nonmagnetic configuration of RuO$_2$.

One might question whether the temperature-invariant XLD signal arises from a Néel temperature ($T_N$) significantly higher than 450 K, the maximum temperature probed in our experiment. Previous studies have reported $T_N$>300 K for RuO$_2$ [20,21], a conclusion widely adopted in subsequent works [8,11,12,14,26]. However, systematic studies on Néel temperature of RuO$_2$ remain scare. The temperature dependence of the XLD signal is expected to follow the scaling relation $XLD(T)\sim|1-T/T_N|^{2\beta}$, where $\beta = 0.26\sim0.4$ for different FM and AFM materials[68,69]. By taking the difference between magnetic and nonmagnetic RuO$_2$ as the magnetic order-related XLD contribution, we simulated the temperature-dependent XLD signal for various assumed $T_N$ values (400 K, 450 K, 500 K and 800 K). As illustrated in Fig. 4(e) and (f), the experimentally observed XLD signal at both Ru M$_3$ edge and O K edge show pronounced deviations from the predicated curves, even when assuming an unrealistically high $T_N$ of 800 K. This discrepancy demonstrates that an antiferromagnetically ordered RuO$_2$ film−regardless of whether its $T_N$ exceeds 450 K or approaches 800 K−should exhibit detectable variations in magnetic-order-related XLD signal across the 310−450 K range. The absence of such temperature dependence provides compelling evidence for the nonmagnetic nature of the RuO$_2$ film.

Accurately defining a universal lower detection limit for XLD is challenging. However, by assuming that the XMLD signal at the Ru M-edge or O K-edge is proportional to the square of the magnetic moment (XMLD $\propto m^2$), we can estimate an upper bound for the ordered moment. Using the benchmark XMLD scaling of ~1.7% at the Ru M$_3$-edge and ~40.4% at the O K-edge for a moment of 0.9 μ$_B$, and considering the RMS noise levels of our XLD signal (~0.21% at Ru M$_3$-edge and ~0.30% at O K-edge), the absence of any discernible signal implies an upper bound for the Ru moment of approximately 0.31 μ$_B$ (derived from the Ru M$_3$-edge data) and 0.077 μ$_B$ (from the O K-edge data).

We note that the size of the x-ray spot used in this XLD measurement is approximately 100 μm by 100μm in ALS and 87 μm by 38μm in SSRF, which may encompass multiple AFM domains, potentially resulting in an averaged, diminished XLD signal. However, space-resolved Photoemission electron microscopy (PEEM) image in Fig. 3 (h) and Supplementary Fig. S7[51] reveal that the RuO$_2$ film exhibits a homogeneous gray scale at both Ru M$_3$ edge and O K edge within the 20-um and 56-um field of view, effectively ruling out the presence of multiple domains.

It is important to exercise caution when considering the effects of sample heating and field cooling of RuO$_2$ films, as those processes could lead to significant alternations in the Ru atomic environment (Supplementary Fig. S8(a)[51]) and substantial oxygen leakage (Supplementary Fig. S8(b)[51]). The similar XAS (Supplementary Fig. S8(c), (d)[51]) and XLD signals [Fig. 3(f)] observed before and after sample heating suggest that these factors were carefully controlled during the experiment.

The striking contrast between the calculated sizable magnetic order-related XLD signal and its near-temperature-invariant, cooling-field-independent XLD signal supports the conclusion that the RuO$_2$ film in our study is nonmagnetic. Our observation reveals a robust crystal-field-derived XLD signal in RuO$_2$ films, with no detectable contribution from magnetic ordering. As demonstrated in Fig. 2(d) and (e), the anisotropic crystal symmetry induces a distinct crystalline orientation-dependent variations in the XLD signal, which is verified by multiple scattering calculation (Part 3 and Supplementary Fig. S4(e)(f)[51]). These findings establish XLD as a powerful probe for characterizing the crystal quality and local atomic environments in altermagnetic materials. We note that our results differ from recent reports [70,71], which attribute the XLD signal exclusively and 80% to magnetic ordering, respectively. This discrepancy persists in different works, highlighting the importance of carefully distinguishing between crystal-field and magnetic-order contributions in XLD measurements.

In summary, we investigated the presence or absence of antiferromagnetic order in epitaxial RuO$_2$ films using a combination of XLD/PEEM measurements and multiple scattering calculations. The near-temperature-invariant and cooling field direction-independent XLD response deviated markedly from the multiple scattering-calculated XLD signals associated with magnetic ordering, providing compelling evidence for the absence of any magnetic order in RuO$_2$. Clear XLD signals were observed at the Ru M edge and the O K edge for different crystal surfaces of RuO$_2$, which are associated to crystal surface-dependent crystal fields. These findings offer new insights into the magnetic properties of $RuO_2$ and its potential applications in spintronic devices.

The project is primarily supported by the Development Program of China (No. 2023YFA1406400), National Natural Science Foundation of China (Grant No. 12174364), the National Key Research, and National Natural Science Foundation of China (12104003, 12241406). This work was partially carried out at the USTC Center for Micro and Nanoscale Research and Fabrication. This research used resources of Beamlines XMCD-A and XMCD-B (Soochow Beamline for Energy Materials) at NSRL. The XMLD measurements were performed at the vector magnet endstation at BL07U at SSRF, beamline 4.0.2 of ALS, and beamline XMCD-a at NSRL. The PEEM measurements were performed at beamline 09U at SSFR and beamline 11.0.1 at ALS. This research used resources of the Advanced Light Source, a DOE Office of Science User Facility under contract no. DE-AC02-05CH11231.

#S. W. and C. W. contributed equally to this work. *contact authors: mmyang@ahu.edu.cn, binhong@buaa.edu.cn, liqian89@ustc.edu.cn.